\documentclass[conference]{IEEEtran}
\IEEEoverridecommandlockouts
\usepackage{cite}
\usepackage{amsmath,amssymb,amsfonts}
\usepackage{algorithmic}
\usepackage{graphicx}
\usepackage{textcomp}
\usepackage{xcolor}

\usepackage{multirow}

\def\BibTeX{{\rm B\kern-.05em{\sc i\kern-.025em b}\kern-.08em
    T\kern-.1667em\lower.7ex\hbox{E}\kern-.125emX}}
\begin{document}

\title{VRCockpit: Mitigating Simulator Sickness in VR Games Using Multiple Egocentric 2D View Frames}

\author{\IEEEauthorblockN{Hao Chen}
\IEEEauthorblockA{\textit{Department of Computing} \\
\textit{Xi'an Jiaotong-Liverpool University}\\
Suzhou, China}
\and
\IEEEauthorblockN{Rongkai Shi}
\IEEEauthorblockA{\textit{Department of Computing} \\
\textit{Xi'an Jiaotong-Liverpool University}\\
Suzhou, China}
\and
\IEEEauthorblockN{Diego Monteiro}
\IEEEauthorblockA{\textit{Digital Engineering School} \\
\textit{ESIEA}\\
Laval, France}
\and
\IEEEauthorblockN{Nilufar Baghaei}
\IEEEauthorblockA{\textit{School of Information Technology and Electrical Engineering} \\
\textit{The University of Queensland}\\
Brisbane, Queensland}
\and
\IEEEauthorblockN{Hai-Ning Liang\textsuperscript{*}}
\IEEEauthorblockA{\textit{Department of Computing} \\
\textit{Xi'an Jiaotong-Liverpool University}\\
Suzhou, China}
\thanks{\textsuperscript{*}Corresponding author (Email: haining.liang@xjtlu.edu.cn)}
}

\maketitle

\begin{abstract}
 Virtual reality head-mounted displays (VR HMDs) have become a popular platform for gaming. However, simulator sickness (SS) is still an impediment to VR's wider adoption, particularly in gaming. It can induce strong discomfort and impair players' immersion, performance, and enjoyment. Researchers have explored techniques to mitigate SS. While these techniques have been shown to help lessen SS, they may not be applicable to games because they  
 %For example, Teleport, which changes continuous movement into discrete jumps, 
 cannot be easily integrated into various types of games without impacting gameplay, immersion, and performance. In this research, we introduce a new SS mitigation technique, VRCockpit. VRCockpit is a visual technique that surrounds the player with four 2D views, one for each cardinal direction, that show 2D copies of the areas of the 3D environment around the player. To study its effectiveness, we conducted two different experiments, one with a car racing game, followed by a first-person shooter game. Our results show that VRCockpit has the potential to mitigate SS and still allows players to have the same level of immersion and gameplay performance.
  \end{abstract}


\section{Introduction}
Virtual reality (VR) has gained mainstream acceptance with the mass commercialization of VR head-mounted displays (HMDs). VR provides users highly immersive experiences, which can make games more attractive and enjoyable. As such, gaming has becomes one of its main uses. Despite its positive affordances, the adoption of VR games still faces some challenges. In the saturated market of games, it is important to grab the players' attention and retain it from the initial times they try and play the game(s) \cite{b1}. This is where VR does not do so well because playing VR games often causes discomforts from simulator sickness (SS), which can happen within the first interactions \cite{b2, b3}. SS can lead to an unpleasant experience and a negative perception of VR~\cite{b45}.

\begin{figure}[tpb]
  \centerline{\includegraphics[width=\columnwidth]{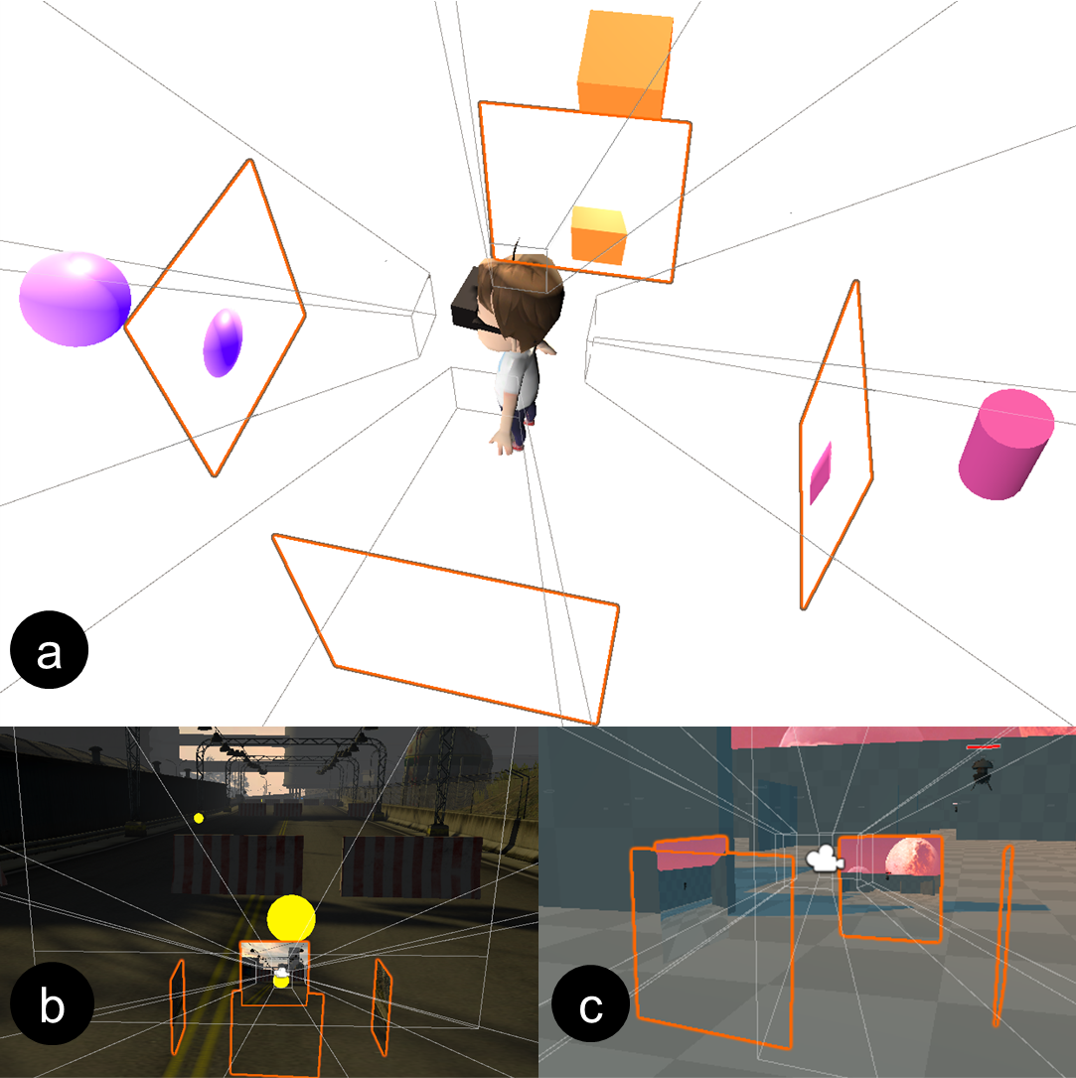}}
  \caption{(a) The conceptual design of \textit{VRCockpit}. The user is surrounded by four planes, which show 2D copies of the view behind them. The users see the 2D views instead of the full 3D renderings behind them. (b and c) Two examples of applying \textit{VRCockpit} in actual games: a car racing game and a first-person shooter game.}
  \label{fig:vrcockpit}
\end{figure}

%To help VR users reduce their SS symptoms, various techniques have been proposed and studied \cite{b4, b5, b6, b7, b8, b9, b10, b11, b12, b13}. %Some of them, most famously \textit{Teleport}, aim to reduce the visual flux by eliminating walking or movement within the virtual environment (VE), often trading off playability in some situations \cite{b14}. In addition, it is not applicable to many types of games, such as car racing games or other sports games, where the main gameplay requires continuous movement of the virtual body. Not all techniques are so limiting. Some successful techniques can reduce SS discomfort by adding additional visual elements to the VE, such as reference frames. Examples include the use of grids in \cite{b6} and even Minions in \cite{b12}. With these techniques, users can navigate more naturally and comfortably but have to rely on extra visual aids in the VE.

To help VR users reduce their SS symptoms, various techniques have been proposed and studied \cite{b4, b5, b6, b7, b8, b9, b10, b11, b12, b13}. Mitigation techniques have been shown to be effective in many non-gaming scenarios, but in games they often show inconsistent or even negative results (e.g., \cite{b15, b16}). Games are complex and visually rich environments and players are required to do frequent (and often rapid) changes of viewpoints (e.g., racing and first-person shooter games \cite{b11,b14,b5,b17,b16,b47, b49}). It is challenging to design techniques that are effective in reducing SS and, at the same time, maintaining a suitable level of immersion and gameplay as good as without such techniques. For example, Teleport has been reported to be efficient in SS reduction, but it makes navigation discrete, which is not suitable for action-based games where continuity of scenes is crucial \cite{b14,b18}. Similarly, using rest frames requires adding additional fixed visual elements to the VE for players to look at continuously. This approach could be limiting, as the player should avoid looking away from these visual elements but should have their visual focus on them frequently. Other common techniques are based on blurring parts of the display \cite{b5, b15, b16} or reducing its field-of-view \cite{b8, b16} and, as such, would likely block away certain parts of the VE, making them inaccessible or invisible to players. Other researchers have explored providing first- vs. third-perspective views \cite{b48,b50,b51}. Some results indicate that third-person VR is less likely to make people have SS compared with first-person, but the former is not perceived as immersive. In addition, some games, like first-person shooters, are more enjoyable and natural when played in first-person view.

This research presents a new visual mitigation technique, \textit{VRCockpit}, for VR games that, as our results show, allows players to have the same level of immersion and performance. VRCockpit surrounds the players with four 2D views, one for each cardinal direction, that show 2D copies of the areas of the 3D environment behind them (see Fig.~\ref{fig:vrcockpit}). It does not impose a non-natural way of navigation, add any additional fixed frames, or block parts of the VE from players' view. We ran a user experiment with a car racing game and a second follow-up experiment with a first-person shooter game. In the two experiments, we measured players' performance, immersion, and usability against their SS levels. Our results show that VRCockpit is a simple yet viable, cost-efficient, and effective technique for SS mitigation in VR racing types of games. 

In this paper, we make the following two contributions: (1) VRCockpit, a new SS mitigation technique for VR games; (2) two user studies with two different game contexts and their analysis of VRCockpit's effect on immersion, gameplay, and performance. Our findings suggest that VRCockpit can be applied to different types of games, can maintain the same level of immersion (compared to not using it), and can lead to the same level of gameplay performance.

This paper is organized as follows: we first introduce our theoretical background and related literature on existing techniques, how they influenced our research, and some of their relevant findings. We then introduce VRCockpit and its design rationale, followed by two studies. We then present a discussion about the results and possible directions for future work. The paper ends with a conclusion. 

\section{Related Work}
\subsection{Simulator Sickness in VR Games}
Simulator sickness (SS) refers to side effects that can occur during or after experiencing VR \cite{b3, b46}. It is common but has not been fully understood so far \cite{b19}. Two of the most widely accepted theories that can potentially explain SS are: the sensory conflict \cite{b20, b21} and the postural instability \cite{b22, b23}. The former attributes SS to a conflict between sensory systems. Specifically, the mismatch between visual inputs and people's vestibular system perception is the main inducement of SS. The postural instability theory posits that a person's inadaptability to the long-term postural instability leads to SS. In addition, the rest frame hypothesis \cite{b24, b25} states that fixed environmental items can be used as references in the human brain for viewing the world. Accordingly, researchers have attempted to use fixed gazing points or other visual objects to reduce SS \cite{b7, b12}. 

\subsection{Mitigation Techniques in Games}
Since SS is always a negative factor hindering the adoption of VR games by users, mitigation techniques have been proposed. Locomotion is a very common and necessary activity for moving in a 3D VE and is considered one of the most significant interaction components of VR experiences \cite{b26}. However, locomotion is also related to the severity of SS because it can elicit either sensory conflict or postural instability \cite{b27, b28}. Accordingly, Teleport has been developed to mitigate SS during locomotion and is now frequently deployed in VR applications \cite{b4}. It allows players to jump to discrete destinations without experiencing continuous transitional movement. Discrete navigation jumps are meant to minimize sensory conflicts and, as such, are thought to be helpful in alleviating SS symptoms. Even if Teleport has shown to be effective and has been applied in some commercial games (i.e., Robo Recall\footnote{https://www.epicgames.com/roborecall/en-US/home}), prior research also indicates that it is detrimental to completing VR tasks due to disorientation \cite{b30, b31, b9}. Furthermore, the technique does not allow continuous movement, and therefore may not be quite appropriate in other game contexts such as car racing, sports, or even first-person shooter games \cite{b14}.

There are other types of visual techniques based on the rest frame hypothesis, which tend to add information to VE. \textit{Gaze Point} \cite{b12} is one example, which offers users a central fixed rest frame. As such, users can focus on the reference point to help mitigate SS. Similarly, Clarke et al. \cite{b7} reported that a target reticule in the center of the screen alleviated dizziness and some SS symptoms. Other types of rest frame techniques have been tested in previous studies to varying degrees of success. Cao et al. \cite{b6} introduced a black see-through mental net as the stationary reference. Their results showed that the net can help reduce SS levels, especially discomfort. Besides, a virtual nose embedded in the middle bottom part of the view has been tested in \cite{b32, b33, b34}, and has been shown to have potential in mitigating SS because it provides a consistent sense of stability. 

Other visual techniques aim to reduce SS by lessening the visual information that users perceive. For instance, viewpoint snapping \cite{b17} and rotation blurring \cite{b5} intentionally reduce visual information the user can receive during rotational movement in VR environments. As such, the illusion of self-motion is also reduced. Likewise, research has found that field of view (FoV) reduction can alleviate SS \cite{b8, b35, b13, b10}. By limiting the display's FoV, the feeling of self-motion may be reduced, which, according to the sensory conflict theory, can assist prevent sensory conflicts and therefore mitigate SS \cite{b43}. Also, dynamic FoV modification \cite{b13} has been proposed to improve issues of lower presence, enjoyment, and user performance. However, such techniques lead to information loss and hence to lower performance in games \cite{b16}.

% A recent approach that combines the properties of several visual techniques, such as Gaze Point and FoV reduction, was proposed in \cite{b11}. It shows benefits above a basic gaze point by displaying an "image" of the 3D virtual world as a 2D view. Their results demonstrated that using 2D frame(s) to display the 3D virtual world contributed to the good outcomes in VR FPS games (an improved performance, a good level of immersion, and reduced level of SS in this type of fast paced game). 

% \subsection{Effect of Mitigation Techniques across Players}
% It is important to note that prior research (e.g., \cite{b5, b35}) has shown that SS mitigation techniques, while having beneficial effects to some users, can cause others to feel greater levels of discomfort than without the technique(s). As such, it is important to analyse the benefits of the technique not only on average but on how effective it is for the most vulnerable users, i.e., those most likely sensitive to SS, because they are the ones most likely to be averse to adopting VR. It is possible that these techniques could have been deemed more successful if presented to the participants who are affected positively by them, without those unaffected by SS and affected negatively by the same technique.

\begin{figure*}[htbp]
    \centerline{\includegraphics[width=\linewidth]{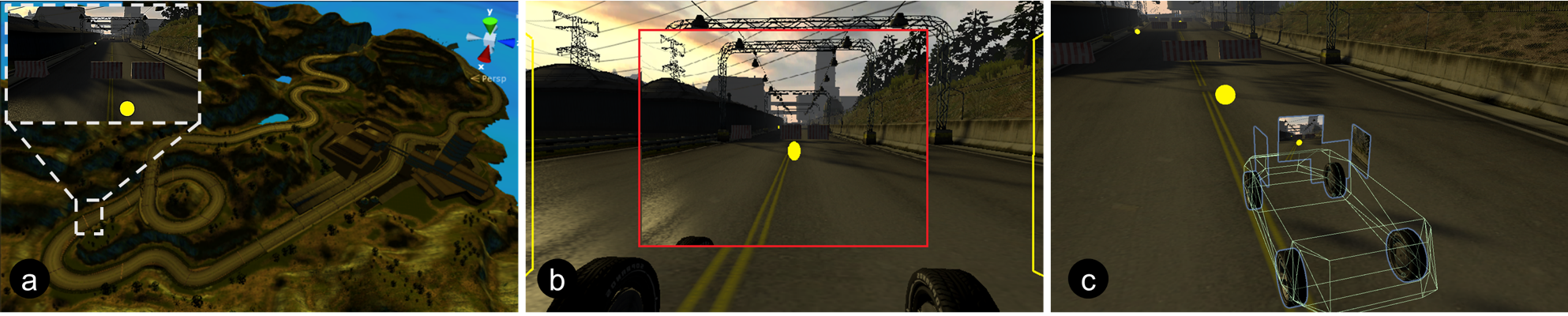}}
    \caption{(a) A bird's view of the whole race track in the game. The zoomed-in screenshot shows a coin to be collected and barriers to be avoided. (b) An example of the first perspective of \textit{VRCockpit} in the game. (c) A third perspective of \textit{VRCockpit} in the game. The car's body was made to be invisible.}
    \label{fig:GameEnv}
\end{figure*}

\section{VRCockpit}
VRCockpit (CP for short) is implemented using render texture in Unity3D, a popular game engine that supports VR devices. CP is composed of four 2D frames, one for each of the cardinal directions around the user (i.e., front, back, left, and right; see Fig.~\ref{fig:vrcockpit}). Each frame displays a copy of the area of the 3D VE behind the frame. They are fixed and cannot be rotated and zoomed in. The borders in the figures in this paper are used to highlight the positions of frames, but they were not shown in the game to avoid separating the 2D views from the 3D VE or adding extra visual elements to it. In each direction, the 2D frame occupies nearly 30\% of users' FoV and all of them are of the same size. This was found to be the ideal size by participants in pilot studies, as users can still see the 3D VE around them between the frames and, at the same, allow the frames to display the main content of interest within them. Fig.~\ref{fig:vrcockpit}b and c show how CP looks in an actual gaming scenario. The four frames are stationary in their positions relative to the user, but their content is updated based on the changes in the areas behind them. As such, it would not change the navigation process. Also, unlike existing techniques that blur or block out part of the screen, CP would not lead to information loss in games but can help reduce SS by transferring 3D content into 2D views.

We also explored other design possibilities, like making the frames move in synchronicity with the user's head. However, feedback from our pilot studies' participants showed that this was unsuitable because they would lose the sense of `3Dness' of the environment and could be somewhat disorientating when experiencing frequent changes. Their suggestion was to keep the frames fixed because they could make rapid eye movements (or head movements) to see through the gaps between the frames and the upper and lower areas not covered by the frames, which was helpful in increasing their immersion in the environment. In addition, fixing the frames' location in relation to the user allowed consistency of views, which has been found to be useful in reducing SS symptoms \cite{b6, b8, b11}.

To evaluate the effects of VRCockpit in reducing SS and players' performance in a fast-paced gaming context, we conducted an experiment with a car racing game as the testbed environment. A second follow-up experiment with another game (a first-person shooter) shows that the technique can be applied to other types of games with similar results (described in Section~\ref{section:follow-up.experiment}).

\section{Experiment}
We compared VRCockpit (CP) with the normal version (Normal) of the game with no technique applied to it. The experiment followed a within-subjects design. All participants played both versions of the car racing game. 

\subsection{Game Environment}
This was an in-house developed VR first perspective car racing game, consisting of turns and steep slopes for players to traverse through (see Fig.~\ref{fig:GameEnv}a). In addition, there were barriers along the way that players needed to avoid and coins to collect, as shown in Fig.~\ref{fig:GameEnv}a. %Three barriers randomly placed
Any uncollected coin would disappear and after the players completed a lap, all coins would be refreshed. Players were required to finish two laps within the same game condition. The car speed was set in the range of 0-70 km/hour (or its equivalent in Unity units). Players were not allowed to stop the car halfway unless they crashed onto barriers. The racetrack does not have any branches. Thus, the players would basically follow the same path. Fig.~\ref{fig:GameEnv}b demonstrates how VRCockpit was implemented in the game. Because the virtual car's body may be an additional rest frame for participants, which would be an undesired confounding factor, the virtual car's body was made invisible (see Fig.~\ref{fig:GameEnv}c).

% This one is a bit too large. 
% \begin{figure*}
%     \centerline{\includegraphics[width=\linewidth]{Fig (Jul 19)/CarRacingGame.PNG}}
%     \caption{Game elements in the car racing game. (a) A bird's view of the whole race track in the game. (b) A screenshot of the game showing a coin to be collected and barriers to be avoided. (c) An example of the first perspective of VRCockpit in the game. (d) A third perspective of VRCockpit in the game.}
%     \label{fig:GameEnv}
% \end{figure*}

\subsection{Evaluation Metrics}
After each condition (CP and Normal), two questionnaires were given to the participants. First, a Simulator Sickness Questionnaire (SSQ) \cite{b40} was adapted to measure the sickness level in terms of different symptoms. Each symptom was rated using a 5-point Likert Scale, with 0 as the least severe and 4 as the most severe. The ratings for each symptom were further grouped into four sub-scores: Nausea, Oculomotor, Disorientation, and Total. Second, an Immersive Experience Questionnaire (IEQ) \cite{b41} with the same rating scale as SSQ was used to elicit participants' perceived immersion in the VR game.

In addition, user performance was determined based on objective measurements. During the experiments, we recorded the total time used (Time), the number of car crashes (Crashes), and coins collected (Coins) to reflect user performance.
% total driving distance (Distance) 

\subsection{Participants and Apparatus}
We recruited 18 participants (3 females, 15 males) whose ages ranged from 18 to 22 ($M=19.63, s.d.=1.50$) from a local university. All the participants reported that they had normal or corrected-to-normal vision and had no history of colour blindness or health issues. Thirteen participants (72.22\%) had some experience with VR systems. Two participants (11.11\%) reported that they felt sick playing PC car racing games.

The virtual environment was rendered via an HTC VIVE Cosmos in this experiment. The VR HMD was connected to a desktop with 16GB RAM, an Intel Core i7-9700k CPU @ 3.60GHz, a GeForce GTX 2080Ti dedicated GPU. Oculus handheld controllers were used to control the movement of the virtual car.%, which is in line with other studies \cite{b16,b44}. 

\subsection{Experimental Procedure}
To eliminate the post-exposure effects of VR sickness, we conducted a two-session experiment where participants only played one version of the game in each session, resting for at least one day. The order of the game version was counterbalanced with a Latin square design to minimize the carry-over effects. At the beginning of the first session, we asked participants to complete a pre-experiment questionnaire to collect demographics and previous gaming experience information, followed by an introduction to the rules and controls of the game. Participants then played the first version of the game, and after this, they were asked to complete the two aforementioned questionnaires based on their experience in the game. The second session was conducted at least 24 hours later to give participants enough rest. Participants played the second version of the game and completed the same questionnaires.

\subsection{Hypotheses}
This experiment aims to verify the following hypotheses: 
\begin{itemize}
    \item \textbf{H1}. Participants will have lower SS in CP than Normal.  
    \item \textbf{H2}. Participants' perceived immersion will not differ significantly between CP and Normal.  
    \item \textbf{H3}. Participants' game performance will not differ significantly between CP and Normal.
\end{itemize}

\section{Results}
We report results of inferential statistics with the support of data visualizations. A Shapiro-Wilk test was first conducted to check the normality of the data. For normally distributed data ($p>.05$), we conducted dependent t-tests. We used Wilcoxon signed-rank tests for the data that was not normally distributed. Spearman's correlation test was used to examine relationships.

\subsection{SSQ} 
\begin{figure}[tbp]
  \centerline{\includegraphics[width=\columnwidth]{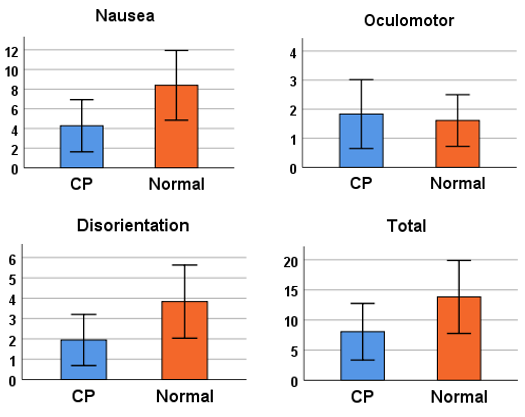}}
  \caption{Mean results of the SSQ data in the experiment according to the four sub-scales. Bars represent 95\% confidence intervals.}
  \label{fig:SSQ}
\end{figure}

Fig.~\ref{fig:SSQ} shows a summary of the results of the SSQ data. All the four sub-scales were not normally distributed, as identified by a Shapiro-Wilk test. After conducting Wilcoxon tests, we found Nausea was significantly lower in CP ($Mdn=1.5$) than in Normal ($Mdn=6.5, Z=-2.103, p=.035$). Disorientation was also significantly lower in CP ($Mdn=0.5$) than in Normal ($Mdn=4.0, Z=-2.196, p=.028$). However, there was no significant difference in Oculomotor ($Z =-.449, p=.653$) and Total ($Z=-1.658, p=.097$).

\subsection{Subjective Immersion}
The descriptive analysis of participants' perceived immersion is summarized in Table.~\ref{tab:upandimm}. Wilcoxon test did not reveal a significant difference between two conditions ($Z=-.711, p=.477$). Spearman's correlation test did not reveal any significant relationships between immersion level and reported SS in two conditions ($p>0.5$ for all cases). 
% Fig.~\ref{fig::IMExpB} shows the data for Immersion between two versions.

% \begin{figure}[htbp]
%   \centerline{\includegraphics[width=0.48\columnwidth]{Fig/Car_Immersion.png}}
%   \caption{Mean results of Immersion in Experiment B. Bars represent 95\% confidence intervals.}
%   \label{fig::IMExpB}
% \end{figure}

\subsection{User Performance}
% Fig.~\ref{fig:UP} shows the data for Coins, Time, and Crashes. 
Table.~\ref{tab:upandimm} summarizes the results for Coins, Time, and Crashes. According to the results of normality test, only Coins was normally distributed. Dependent t-tests did not reveal significant difference in Coins ($t(17)=-1.042, p=.312$). In addition, Wilcoxon tests showed no significant difference in Crashes ($Z=-1.481, p=.138$) and Time ($Z=-.022, p=.983$). 

We performed Spearman's correlation test to investigate the relationships between participants' user performance and reported SS, and between user performance and perceived immersion. No significant correlations were identified in the CP condition. While in the Normal condition, there were significant negative relationships between Crashes and Nausea ($r_s=-.659, p=.003$), Crashes and Disorientation ($r_s=-.658, p=.003$), Crashes and Total ($r_s=-.629, p=.005$).

% \begin{figure}[tbp]
%   \centerline{\includegraphics[width=\columnwidth]{Figure/UP.png}}
%   \caption{Mean results of Coin, Time, and Crashes in the experiment. Bars represent 95\% confidence intervals.}
%   \label{fig:UP}
% \end{figure}

\begin{table}[htbp]
  \caption{A summary of the means, standard deviations, and medians of participants' perceived immersion and user performance (represent using $M, s.d., Mdn$, respectively).}
  \begin{center}
  \begin{tabular}{|c|c|c|c|c|}
  \hline
  \textbf{Measure} & \textbf{Condition} & \textbf{\textit{M}} & \textbf{\textit{s.d.}} & \textbf{\textit{Mdn}}\\ \hline
  \multirow{2}{*}{Immersion} & CP & 25.22 & 7.49 & 24.50\\ \cline{2-5}
   & Normal & 27.39 & 8.70 & 27.50\\ \hline
  \multicolumn{5}{|c|}{  } \\ \hline 
  \multirow{2}{*}{Time} & CP & 543.00 & 110.70 & 531.26\\ \cline{2-5}
   & Normal & 556.36 & 153.47 & 502.30\\ \hline
  \multirow{2}{*}{Crashes} & CP & 14.28 & 12.55 & 11.00\\ \cline{2-5}
   & Normal & 9.22 & 7.60 & 7.00 \\ \hline
  \multirow{2}{*}{Coins} & CP & 45.67 & 13.82 & 47.00\\ \cline{2-5}
   & Normal & 48.61 & 15.75 & 50.00\\ \hline
  \end{tabular}
  \end{center}
  \label{tab:upandimm}
\end{table}

\section{Discussion}
The results show that VRCockpit led to significantly lower symptoms in Nausea and Disorientation for all participants than without it, indicating that the technique helped typical players feel lower SS in the car racing game. This finding confirms our hypothesis \textbf{H1}. Moreover, most participants felt either better or indifferent to the technique and most of the participants who felt better had considerable improvements. These participants were the ones who scored the highest when not using any form of mitigation technique (i.e., in the Normal condition). %[I didn't get the point here from the old paper.]} 

There was no significant difference in participants' perceived immersion. In addition, there was no significant difference in user performance in terms of total time used, the number of car crashes, and coins collected in our racing game. These results indicate that VRCockpit would be less likely to affect immersion and users' gameplay negatively. As such, Hypotheses \textbf{H2} and \textbf{H3} are both confirmed. Based on participants' feedback, VRCockpit worked well in the car racing game because it served as car windows and was closer to real driving situations. A natural combination between the technique and the virtual environment and scenario appears to be helpful for how users perceive the technique. 

Interestingly, we found that the number of crashes and reported SS symptoms were negatively correlated in the car racing game where no SS mitigation technique was applied. During gameplay, participants drove at similar speeds and were not required to make any stops, except when crashing onto the barriers. When participants drove smoothly, the surroundings changed rapidly, which led to the accumulation of discomfort. While such accumulation was interrupted when a crash happened. Participants had a short rest (usually a few seconds) and this helped alleviate some symptoms. However, these negative correlations were not found when VRCockpit was used in the game. The main reason was that the technique had reduced the SS symptoms, so the effects of crashes were not significant. 

In summary, the results from our experiment suggest that VRCockpit in a car racing game not only had an effect on mitigating SS but did not affect user performance and immersion---this is a positive indication for VRCockpit's use in gaming environments that require continuous movement. The car racing game can be a typical game involving virtual navigation. In other words, players are mainly focusing on the motion control in the VEs during the gameplay. To validate the usefulness of VRCockpit in a more complex game scenario, we ran a follow-up experiment using a First-Person Shooting (FPS) game. An FPS game is generally more dynamic compared to a car racing game, because it involves weapon-based fighting in addition to virtual navigation. Thus, we used an FPS game as an additional testbed and proof-of-concept of VRCockpit.

\begin{figure}[tbp]
  \centerline{\includegraphics[width=\columnwidth]{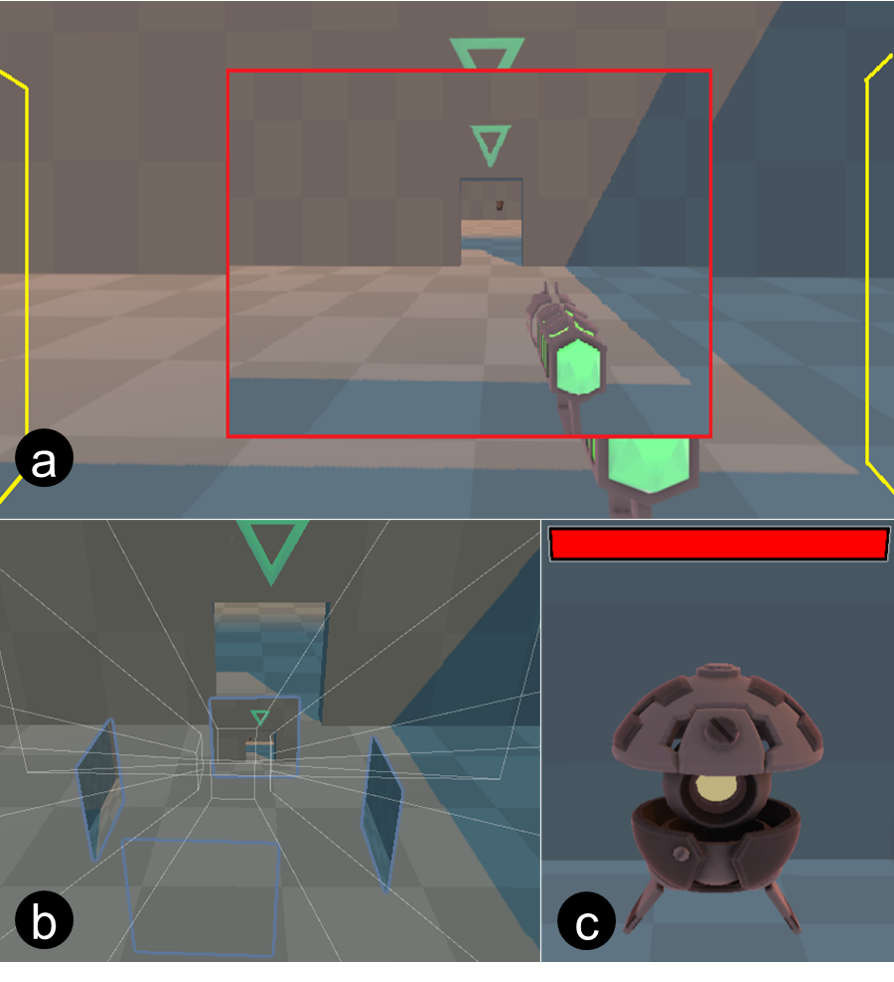}}
  \caption{Game elements in the FPS game. (a) A first perspective view of VRCockpit in the game. The frames with yellow lines show the left and right views–but were not actually rendered in the application, like in the car racing game. There was another 2D frame behind the user. (b) A third perspective of VRCockpit in the game. (c) A picture of a robot enemy that the players need to seek and destroy.}
  \label{fig:FPSGameEnv}
\end{figure}

\section{Follow-Up Experiment} \label{section:follow-up.experiment}
This follow-up experiment sought to further investigate whether VRCockpit can help players reduce the Simulator Sickness symptoms without affecting their performance and immersion in a more complex and dynamic VR game scenario---an FPS game. 

\begin{table*}
  \caption{A summary of descriptive data and inferential results of participants' perceived sickness, perceived immersion, and user performance ($M, s.d., Mdn$ stand for mean, standard deviation, and median, respectively).}
  \begin{center}
  \begin{tabular}{|c|c|c|c|c|c|}
  \hline
  \textbf{Measure} & \textbf{Condition} & \textbf{\textit{M}} & \textbf{\textit{s.d.}} & \textbf{\textit{Mdn}} & \textbf{Wilcoxon tests}\\ \hline
    \multirow{2}{*}{Nausea} & CP & 0.40 & 0.97 & 0.00 & \multirow{2}{*}{$Z=-2.201, p=.028$}\\ \cline{2-5}
   & Normal & 5.00 & 7.54 & 1.50 & \\ \hline
    \multirow{2}{*}{Oculomotor} & CP & 0.70 & 0.82 & 0.50 & \multirow{2}{*}{$Z=-.108, p=.914$}\\ \cline{2-5}
   & Normal & 1.10 & 1.85 & 0.50 & \\ \hline
    \multirow{2}{*}{Disorientation} & CP & 0.00 & 0.00 & 0.00 & \multirow{2}{*}{$Z=-2.226, p=.026$}\\ \cline{2-5}
   & Normal & 2.40 & 3.86 & 1.00 & \\ \hline
   \multirow{2}{*}{Total} & CP & 1.10 & 1.29 & 1.00 & \multirow{2}{*}{$Z=-1.992, p=.046$}\\ \cline{2-5}
   & Normal & 8.50 & 12.93 & 3.00 & \\ \hline
  \multicolumn{6}{|c|}{  } \\ \hline
    \multirow{2}{*}{Immersion} & CP & 23.20 & 11.12 & 23.50 & \multirow{2}{*}{$Z=-1.123, p=.262$}\\ \cline{2-5}
   & Normal & 20.90 & 8.02 & 20.50 & \\ \hline
  \multicolumn{6}{|c|}{  } \\ \hline 
    \multirow{2}{*}{Time} & CP & 322.57 & 77.03 & 311.44 & \multirow{2}{*}{$Z=-1.070, p=.285$}\\ \cline{2-5}
   & Normal & 346.85 & 113.13 & 309.32 & \\ \hline
    \multirow{2}{*}{Distance} & CP & 1861.15 & 456.30 & 1884.28 & \multirow{2}{*}{$Z=-.764, p=.445$}\\ \cline{2-5}
   & Normal & 1956.96 & 500.71 & 1775.75 & \\ \hline
    \multirow{2}{*}{Shots \newline Received} & CP & 197.40 & 99.84 & 153.50 & \multirow{2}{*}{$Z=-.051, p=.959$}\\ \cline{2-5}
   & Normal & 201.90 & 122.38 & 162.50 & \\ \hline
  \end{tabular}
  \end{center}
  \label{tab:FPSresults}
\end{table*}

\subsection{Experimental Setup}
The FPS game was built based on the FPS Microgame Template\footnote{https://learn.unity.com/project/fps-template} from Unity Learn. Players needed to navigate quickly and defeat all the enemy robots in the VE while avoiding getting hit by them to finish the game. Fig.~\ref{fig:FPSGameEnv}a and b show how VRCockpit was implemented in the FPS game. The game environment was made up of towering grey walls with no additional visual clues to help with path memory, while there were no confusing or hidden pathways. Furthermore, the maximum ammo of the gun was 5 and it would reload once depleted. This mechanic was designed to motivate players to move frequently and balance shooting and hiding. The adversaries were same-sized robots (see Fig.~\ref{fig:FPSGameEnv}c), which would follow and attack the players once the players were within their range of detection. Each robot could withstand up to 14 bullet shots from the players.

The experimental design and procedure were similar as in the first experiment. Specifically, participants needed to complete a two-session experiment with a normal version of the game (Normal) and a version that incorporated VRCockpit (CP). We used the same SSQ and IEQ for measuring the sickness and immersion level. For participants' performance, we recorded the total time that participants spent in the game (Time), total distance travelled (Distance), and total hits received from robot enemies (Shots Received). We used the same analysis methods and approach as the first experiment. 

We recruited 10 participants (6 females, 4 males) in the follow-up experiment. They were aged between 19 to 28 ($M=21.20, s.d.=2.48$). They all had a normal or normal-to-corrected vision. None of them had any history of colour blindness or other health issues. Seven participants had some experience playing or interacting with VR HMDs. Six reported that they felt discomfort when playing FPS games in a traditional desktop monitor in the past. An Oculus Rift S VR HMD was used to display the virtual environment. Participants used mouse and keyboard to control their movements and shooting behaviors as they lead to better performance in VR FPS games \cite{b42}. 

\subsection{Results and Discussion}
Table.~\ref{tab:FPSresults} summarizes all the results of this second experiment. As shown in the table, Nausea, Disorientation, and Total were significantly lower in CP than in Normal. No other significant differences were found. These results were in line with our results in the first experiment. Given that we had a lower number of participants in the follow-up experiment, we could only provisionally conclude that VRCockpit can also help to mitigate SS symptoms in a VR FPS game without lowering players' immersion and gameplay. However, the follow-up experiment and the results we observed can still be regarded as a proof-of-concept providing further evidence that the technique can be potentially applied to different types of VR games with various difficulty levels and requiring different types of navigation strategies. 

Out of our expectations, although the FPS game is more complex and dynamic compared to the car racing game, participants generally gave lower ratings of the SS symptoms in the FPS game (which can be reflected in the results shown in Fig.~\ref{fig:SSQ} and Table.~\ref{tab:FPSresults}). One possible reason is that, unlike the car racing game, players needed to show more awareness of all sides around them and react quickly if an enemy showed up. In addition, players may be able to stop, move at different speeds in the FPS game, and could find some chances to hide somewhere and wait for reloading, which provided users with some breaks. % However, these comparisons and analyses were out of the scope of this paper and requires future work with rigorously controlled experiment to validate. 

Few participants felt it was somewhat unusual and abrupt to have a set of 2D frames surrounding them. One way that the technique could be better integrated into the FPS is to tell participants that they are wearing a virtual helmet with see-through displays. This may help improve their perception of the technique. As mentioned before, the combination between the technique and the virtual environment and scenario should be natural to improve the game experience. 

\section{Limitations and Future Work}
Because of different design elements between experiments (e.g., time, controller), we were unable to conduct a between-subjects analysis to explore if game types could influence the technique's effectiveness. These choices were by design and guided by prior research. For example, we opted for the typical dual-hand controller for the VR racing game to be in line with other studies (e.g., \cite{b16,b44}). On the other hand, we used a mouse and keyboard in the FPS game, because \cite{b11} observed advantages in using the mouse and keyboard over a dual hand-based controller in a VR FPS game. Also, it is often the case that time and distance are within a similar range for all participants in car racing games, whereas this may not be the case for FPS games, which could afford a broader range in how each participant could behave during gameplay. Given that VRCockpit performed better in the car racing game, it is plausible that our technique is more suitable when applied in this type of continuous fast-paced context. Further explorations like comparing different speeds, acceleration, and size of the 2D frames will be helpful in assessing its effectiveness and application in other scenarios. In this research, we wanted to validate if VRCockpit is effective in mitigating SS while maintaining suitable performance and immersion levels, which we have achieved to a large extent. Optimization issues are part of our future plans.

We can also conduct a comparative study to contrast our technique with other existing SS mitigation techniques involving a larger number of participants. However, such studies are often non-trivial, as direct comparisons are often limited to the original choice of environment and not easily transferable to other environments because it is generally challenging to replicate the technique to other scenarios outside of the original testbed setting. However, as demonstrated in our results, the trade-offs of the mitigation technique can be game-specific. Finding the best matches between technique(s) and game type(s) represents potentially a rich avenue for future research.

\section{Conclusion}
In this paper, we explored ways to mitigate simulator sickness (SS) in virtual reality (VR) games. We proposed a new visual mitigation technique for head-mounted displays (HMDs), VRCockpit. It surrounds users with four egocentric frames, one for each cardinal direction, that display 2D replicas of the 3D environment behind each frame. We conducted two user studies to test our technique with two different games (a car racing game and a first-person shooter game) and compared it with the baseline condition (with no visual mitigation technique). Our results show that VRCockpit is able to reduce SS for participants in general and still allows them to attain similar levels of immersion and performance in the two games. Our findings show that VRCockpit is a useful SS mitigation technique and does not change the navigation process, add additional visual elements to the gaming environment, or blur/blackout portions of the display.

\section*{Acknowledgment}
The authors thank the participants who joined the study and the reviewers for their insightful comments and helpful suggestions that helped to improve our paper. This work was supported in part
by Xi’an Jiaotong-Liverpool University–Key Special Fund (\#KSF-A-03) and the Future Network Scientific Research Fund (\#FNSRFP-2021-YB-41).
% The preferred spelling of the word ``acknowledgment'' in America is without an ``e'' after the ``g''. Avoid the stilted expression ``one of us (R. B. G.) thanks $\ldots$''. Instead, try ``R. B. G. thanks$\ldots$''. Put sponsor acknowledgments in the unnumbered footnote on the first page.

% \section*{References}

% \vspace{12pt}
\end{document}